\documentclass[twocolumn,english,aps,prd,reprint,floatfix,notitlepage,footinbib,preprintnumbers,superscriptaddress,longbibliography]{revtex4-1}
\pdfoutput=1
\usepackage{lmodern}

\usepackage[T1]{fontenc}
\usepackage[latin9]{inputenc}
\usepackage{geometry}
\usepackage{subfigure,lmodern, amsmath,amssymb, graphicx, pifont, adjustbox, bm, xcolor}
\usepackage{amsfonts}
\usepackage{enumitem}
\usepackage{comment}
\usepackage{mathtools}
\usepackage{float}
\usepackage{slashed}
\usepackage{ragged2e}
\usepackage{array}
\usepackage{bbm}
\usepackage{balance}

\usepackage{nameref}

\usepackage{hhline}

\makeatletter\g@addto@macro\bfseries{\boldmath}\makeatother

\makeatletter\newcommand{\labeltext}[2]{%
  \def\@currentlabel{#1}%
  \label{#2}%
}
\makeatother

\usepackage{stackengine}
\usepackage{esint}
\usepackage[unicode=true,pdfusetitle,
 bookmarks=true,bookmarksnumbered=false,bookmarksopen=false,
 breaklinks=false,pdfborder={0 0 1},backref=false,colorlinks=true]
 {hyperref}
\hypersetup{
 pdfauthor={Gary T. Horowitz, Maciej Kolanowski, Grant N. Remmen, Jorge E. Santos},
 citecolor=black,linkcolor=black,urlcolor=black}

\newcommand{\appendixref}[1]{\hyperref[#1]{appendix~\ref{#1}}}
\def\equationautorefname~#1\null{eq.\,(#1)\null}
\usepackage{breakurl}
\usepackage[hang,flushmargin]{footmisc} 
\allowdisplaybreaks
\makeatletter

\usepackage{etoolbox}
\apptocmd{\thebibliography}{\justifying\setlength{\leftskip}{7.4mm}}{}{} 
 
 \usepackage{relsize}
\usepackage{babel}

\makeatletter
\def\simgt{\mathrel{\lower2.5pt\vbox{\lineskip=0pt\baselineskip=0pt
           \hbox{$>$}\hbox{$\sim$}}}}
\def\simlt{\mathrel{\lower2.5pt\vbox{\lineskip=0pt\baselineskip=0pt
           \hbox{$<$}\hbox{$\sim$}}}}
\makeatother

\usepackage{changepage}

\newcommand{\be}{\begin{equation}}
\newcommand{\ee}{\end{equation}}
\newcommand{\bea}{\begin{eqnarray}}
\newcommand{\eea}{\end{eqnarray}}

\newcommand{\mysec}[1]{\noindent {\bf #1.}---}

\newcolumntype{P}[1]{>{\centering\arraybackslash}p{#1}}

\usepackage{fix-cm}

\definecolor{dartmouthgreen}{rgb}{0.05, 0.5, 0.06}

\begin{document}

\title{Axions Create Singularities on Extremal Horizons}

\author{Gary~T.~Horowitz}
\affiliation{Department of Physics, University of California, Santa Barbara, CA 93106, USA}

\author{Maciej~Kolanowski}
\affiliation{Department of Physics, University of California, Santa Barbara, CA 93106, USA}

\author{Grant~N.~Remmen}
\affiliation{\scalebox{1}{Center for Cosmology and Particle Physics, Department of Physics, New York University, New York, NY 10003, USA}} 

\author{Jorge~E.~Santos}
\affiliation{DAMTP, Centre for Mathematical Sciences, University of Cambridge, Wilberforce Road,
Cambridge CB3 0WA, UK}

\begin{abstract}
\noindent We show that axions cause extremal black holes to have singular horizons. This is true for almost all values of the axion mass and coupling
provided the black hole is rotating and has some arbitrarily small nonzero charge. When the axion mass becomes large, these singularities are related to the recently discovered singularities induced by higher-derivative corrections to the Einstein-Maxwell equations.
Away from extremality, this effect produces anomalously large tidal forces in the vicinity of near-extremal horizons, causing breakdown of the effective theory.

\end{abstract}

\maketitle

\mysec{Introduction}In general relativity, the dynamics of spacetime on macroscopic length scales are generically expected to be insensitive to very massive particles in the spectrum.
This intuition is central to the Wilsonian conception of effective field theory (EFT), in which physics in the infrared decouples from physics in the ultraviolet.
In this paper, we will exhibit a dramatic violation of this expectation, in which massive particles lead to the generation of singularities on the horizons of macroscopic extremal black holes.

Within the Einstein-Maxwell EFT, it was recently shown that $F^4$ corrections---such as are generated at low energies in the standard model---become very important in the vicinity of an extremal black hole \cite{Horowitz:2023xyl, Horowitz:2024dch}, where they lead to singularities (and parametrically large tidal forces for near-extremal black holes)~\footnote{There is another class of important corrections for near-extremal black holes, called Schwarzian modes, that significantly alter their thermodynamics~\cite{Iliesiu:2020qvm, Rakic:2023vhv, Kapec:2023ruw, Kolanowski:2024zrq} and absorption properties \cite{Emparan:2025qqf, Emparan:2025sao, Biggs:2025nzs}. However, as explained in Ref.~\cite{Horowitz:2024dch}, one needs to be orders of magnitude closer to extremality for them to become relevant. Thus, we will happily ignore them in the rest of this paper.}.
In the EFT, these $F^4$ terms are suppressed by inverse powers of an ultraviolet cutoff $m$, signaling the scale of new particles that were integrated out. One might hope that the horizon singularities generated in the EFT are simply artifacts of the derivative expansion and would be removed in a more ultraviolet-complete theory.  Our results show that this is not the case. In fact, we find that the singularity can be stronger with the ultraviolet degree of freedom included.

We consider a massive axion coupled to the Maxwell field through an $F\wedge F$ term. 
The classic QCD axion is famously motivated by the strong CP problem~\cite{Peccei:1977hh,Weinberg:1977ma,Wilczek:1977pj}, and various generalized axion theories~\cite{Kim:1979if,Shifman:1979if,Dine:1981rt,Zhitnitsky:1980tq} are viewed as one of the main dark matter candidates~\cite{Preskill:1982cy,Kim:2008hd,Chadha-Day:2021szb},  with the coupling set by the axion decay constant.
The QCD axion generically couples~\footnote{Specifically, the QCD axion mixes with the neutral pion, and the latter experiences a coupling to $F\wedge F$ via a Wess-Zumino-Witten anomaly due to chiral symmetry breaking~\cite{Wess:1971yu,Witten:1983tw}.} to the electromagnetic sector as well~\cite{Sikivie:1983ip,Wilczek:1987mv}, and this coupling is also an expected feature of string compactifications, where the four-dimensional theory has an abundance of axion-like particles~\cite{Svrcek:2006yi,Arvanitaki:2009fg}. The Einstein-Maxwell-axion theory we consider in this paper is therefore well motivated both from the viewpoint of phenomenology and string theory. 

Within this theory, we study axisymmetric rotating and charged black hole solutions.  
Due to the axion's nonminimal coupling to the photon, the solutions necessarily have axion hair. 
At fixed black hole mass and charge (and fixed axion parameters), solutions exist only for angular momentum $J$ smaller than a maximal value $J_{\max}$.  We start our investigations by constructing the near-horizon geometries (NHGs) of the extremal solutions with $J = J_{\max}$. We find that for sufficiently small $J/(GQ^2)$ and at sufficiently large couplings, these geometries are non-unique. Only one of them survives in the astrophysically relevant limit $J \gg G Q^2$. Following the philosophy of Refs.~\cite{Horowitz:2023xyl, Horowitz:2024dch}, we look for the lowest scaling dimension $\gamma$ of perturbations to this NHG that are induced by connecting to the asymptotically-flat region.   We find that $\gamma < \gamma_0$ where $\gamma_0 \approx 1.6$. Since $\gamma$ is almost never an integer, this strongly suggests that the extremal asymptotically-flat solutions have singular tidal forces at the horizon, and non-extremal horizons would have tidal forces growing with the black hole temperature as $\gamma(\gamma-1)T^{\gamma-2}$.
We confirm this expectation  by constructing the full asymptotically-flat solutions at low temperature.

As in earlier studies \cite{Horowitz:2023xyl, Horowitz:2024dch}, all scalar quantities constructed from the curvature remain finite at the extremal horizon. 
The singularity appears as divergent tidal forces, which if $\gamma<1$ are accompanied by associated divergent electric fields and/or axion gradients.
Despite their non-scalar nature, these are physical singularities that would be felt by an infalling observer. 

One might worry that our results could be spoiled by the superradiant instability that is known to occur for axions around a rotating black hole~\cite{Arvanitaki:2010sy}. However, the axisymmetric sector that we consider is not subject to the usual superradiant instability. 
 Nevertheless, it leaves open the question of whether our near-extremal solutions could be formed dynamically.

While Ref.~\cite{Chen:2024sgx} found that a massive dilaton could generate singular behavior on the horizon of extremal charged black holes, an important difference with the present work is that they required the non-spherical  modes generating the singularity to be externally sourced.
In contrast, for axions the extremal charged rotating black hole solutions {\it themselves} are singular, without any sourced perturbation required~\footnote{Moreover, while the dilaton couples to $F^2$, and while singular dilaton behavior on the horizon is not too surprising in light of Refs.~\cite{Garfinkle:1990qj,Porfyriadis:2023qqo}, the axion couples to $F\wedge F$, which does not even generate a stress tensor, making the singular behavior all the more surprising.}.

\medskip

\mysec{Theory and Methodology}We consider the Einstein-Maxwell-axion theory defined by the Lagrangian density
\begin{equation}
\hspace{-1.5mm}\mathcal{L}\,{=}\,\frac{1}{2\kappa^2}\!\left[
R {-} F_{\mu\nu}F^{\mu\nu} 
{-} 2(\nabla a)^2
{-}2m^2 a^2
{-} g\, a\, F_{\mu\nu}\widetilde{F}^{\mu\nu}
\right]\!.\hspace{-1mm}
\label{eq:lagrangian}
\end{equation}
Here the gravitational coupling is $\kappa^2=8\pi G$, $m$ is the axion mass, and $g$ denotes its coupling to the electromagnetic field.
The Hodge dual of the field strength $\widetilde{F}=\star F$ is written in components as $\widetilde{F}^{\mu\nu}= \epsilon^{\mu\nu\rho\sigma}F_{\rho\sigma}/2$~\footnote{The Levi-Civita tensor is normalized such that $\sqrt{-g}\,\epsilon^{tr\theta\phi}=1$. We work in mostly-plus metric signature, with sign conventions $R_{\mu\nu}=R^\rho_{\;\;\mu\rho\nu}$ and $R^\mu_{\;\;\nu\rho\sigma} = \partial_\rho \Gamma^\mu_{\nu\sigma} -\partial_\sigma \Gamma^\mu_{\nu\rho} + \Gamma^\mu_{\rho\alpha}\Gamma^\alpha_{\nu\sigma} - \Gamma^\mu_{\sigma\alpha}\Gamma^\alpha_{\nu\rho}$. }.
The corresponding equations of motion are
\begin{equation}
\begin{aligned}
E_{\mu\nu}\equiv&\, R_{\mu\nu}-\frac{1}{2}g_{\mu\nu}R\\&-2F_{\mu}^{\phantom{\mu}\rho}F_{\nu\rho}+\frac{1}{2}g_{\mu\nu}F_{\rho \sigma}F^{\rho\sigma}\\&-2\nabla_\mu a\nabla_\nu a+g_{\mu\nu}\nabla_\rho a\nabla^\rho a+m^2a^2 g_{\mu\nu}=0,\label{eq:eomein}
\end{aligned}
\end{equation}
\begin{equation}
E_{\mu}\equiv \nabla^\nu \bar{F}_{\nu\mu}=0,\label{eq:eomA}
\end{equation}
and
\begin{equation}
\nabla_\mu \nabla^\mu a-m^2a-\frac{g}{4}\widetilde{F}_{\mu\nu}F^{\mu\nu}=0,\label{eq:eomphi}
\end{equation}
where $\bar{F}\equiv F+g a \widetilde{F}$.

We search for black hole spacetimes with spherical horizon topology. As such, we work in coordinates $(t,r,\theta,\phi)$ adapted to stationarity and axisymmetry, 
where surfaces of constant $t$ and $r$ are topological two-spheres $S_{tr}^2$.
The black hole horizon is defined as the radius $r=r_+$ at which, below extremality, $g^{rr}$ vanishes linearly. 
The spacetime admits a timelike Killing vector $k\equiv \partial_t$ and an axial Killing vector $\xi\equiv \partial_\phi$, with identification $\phi\sim\phi+2\pi$. 

The most general field ansatz consistent with the above symmetries and the discrete
transformation $(t,\phi)\to(-t,-\phi)$~\footnote{Invariance under  $(t,\phi)\to(-t,-\phi)$ follows from parity arguments upon imposing that the axion transforms as a pseudoscalar, cf. arguments of the Carter-Papapetrou type~\cite{Papapetrou1966,Carter1968,Carter1971,Xie:2021bur}.} is
\begin{equation}
\begin{aligned}
{\rm d}s^2=\,&-Q_1(r,\theta){\rm d}t^2+\frac{{\rm d}r^2}{Q_2(r,\theta)}\\&+Q_8(r,\theta)\left[{\rm d}\theta+Q_9(r,\theta){\rm d}r\right]^2 \\& +Q_3(r,\theta)\left[{\rm d}\phi-Q_4(r,\theta){\rm d}t\right]^2
\\
A=\,&Q_7(r,\theta){\rm d}t+Q_6(r,\theta)\left[{\rm d}\phi-Q_4(r,\theta){\rm d}t\right]\vphantom{\biggl(}
\\
a=\,&Q_5(r,\theta).
\end{aligned}
\end{equation}
To construct finite-temperature solutions, we work in the so-called conformal gauge where
$Q_9(r,\theta)=0$ and $Q_8(r,\theta)$ is related to $Q_2(r,\theta)$. Details of the gauge choice are provided in Sec.~\hyperref[sup:2]{A} of the
Supplemental Material. To proceed, we use numerical methods presented in Ref.~\cite{Dias:2015nua} to determine all the remaining $Q_{1,\ldots,7}$. 
The event horizon $\mathcal{H}^+$ is a null hypersurface located at $r = r_+$, where $Q_1$ vanishes linearly while $Q_4$ approaches the constant value $\Omega_H$, as required by the rigidity theorems~\cite{Hawking:1971vc,Hawking:1973uf,Hollands:2006rj}. The constant $\Omega_H$ is identified with the black hole angular velocity, and the horizon is a Killing horizon generated by the Killing vector $\chi \equiv k + \Omega_H\,\xi$.

One can use stationarity and axisymmetry, together with the equations of motion, to show that the following two Komar-type integrals are gauge invariant and conserved on any $S_{tr}^2$ two-sphere:
\begin{equation}
\begin{aligned}
Q &= \frac{2}{\kappa^2}\int_{S^2_{tr}} \star \bar{F}  ,
\\
J &= \frac{1}{2\kappa^2}\int_{S^2_{tr}}
\left\{
\star\!\left[\mathrm{d}\xi^\flat + 2\,(\iota_\xi A) \bar{F}\right]
- 2\,(\iota_\xi \bar{F})\wedge A
\right\}.
\end{aligned}
\end{equation}
Here $\iota$ denotes the standard interior product, and we have defined $\xi^\flat \equiv g_{\mu\nu} \xi^\nu \mathrm{d}x^\mu$. Taking $S^2_{tr}$ to spatial infinity, $Q$ and $J$ reduce to the electric charge and angular momentum, respectively. To any Killing horizon, one may associate a Hawking temperature and a chemical potential~\cite{Hawking:1974rv},
\begin{equation}
\begin{aligned}
T&=\frac{1}{4\pi}
\sqrt{
\left.\partial_r Q_1\right|_{r=r_+}
\left.\partial_r Q_2\right|_{r=r_+}
}
\\
\mu &=
\left.\iota_{\chi} A\right|_{r\to\infty}-\left.\iota_{\chi} A\right|_{r=r_+},
\end{aligned}
\end{equation}
respectively. By the rigidity theorems, both the temperature and the chemical potential are constant on $\mathcal{H}^+$.

Solutions are fully characterized by four dimensionless parameters,
$\{m G M,\; Q/M,\; J/(G M^2),\; g\}$, where $M$ is the ADM energy of the spacetime. For fixed values of these parameters, however, multiple discrete solutions may exist. 
We find that, at fixed $\{mGM,\, Q/M,\, g\}$, the dimensionless spin
$|J/(G M^2)|$ is bounded by a maximal value $J_{\max}$, beyond which no
solutions exist. Solutions saturating this bound are extremal and have vanishing
Hawking temperature, with $\chi^2$ and $g^{rr}$ vanishing quadratically at $r=r_+$, indicating a degenerate, extremal horizon~\footnote{Not all extremal solutions, defined in this sense, have vanishing Hawking temperature; see, e.g., Ref.~\cite{Dias:2021vve}.}.

For extremal solutions, a scaling (near-horizon) limit can be taken by
introducing coordinates $(\tau,\rho,\theta,\psi)$ defined through
\begin{equation}
t=t_0\,\tau/\lambda, \;\;\;
r=r_+(1+\lambda \rho), \;\;\;
\psi=\phi-\Omega_H t ,
\end{equation}
with $t_0$ a real constant, followed by the limit $\lambda \to 0$. In this limit, the geometry develops an
enhanced $SL(2,\mathbb{R})\times U(1)$ symmetry, characteristic of extremal
black hole NHGs~\cite{Bardeen:1999px,Hartman:2008pb,Kunduri:2013gce,Hadar:2020kry,Porfyriadis:2021psx,Charalambous:2022rre,Porfyriadis:2021zfb,Banerjee:2024zix}, and can be written as
\begin{equation}\hspace{-1mm}
\begin{aligned}
{\rm d}s^2_{\rm NH}&=2r_+^2\Omega^2(\theta)\bigg[-\rho^2{\rm d}\tau^2+\frac{{\rm d}\rho^2}{\rho^2}+\Gamma^2_{\rm NH}\rm d \theta^2\\&\qquad\qquad\qquad +\Lambda^2(\theta)\sin^2\theta({\rm d}\psi+\omega_{\rm NH}\,\rho\,{\rm d}\tau)^2\bigg]
\\
A_{\rm NH}&=r_+[Q_{\rm NH}\,\rho\,{\rm d}\tau\\&\qquad\;\; +a_{\psi}(\theta)\Lambda(\theta)\sin^2\theta({\rm d}\psi+\omega_{\rm NH}\,\rho\,{\rm d}\tau)]
\\
a_{\rm NH}&=\varphi(\theta),
\end{aligned}
\label{eqs:nearh}\hspace{-1mm}
\end{equation}
with $Q_{\rm NH}$, $\omega_{\rm NH}$, and $\Gamma_{\rm NH}$ constants and $\Omega$, $\Lambda$, $a_\psi$, and $\varphi$ some functions of the polar angle. (A specific choice of $t_0$ has been made so that the first term in the line element takes the form given above.) Both $Q$ and $J$ can be expressed purely in terms of the NHG and may therefore be matched to the asymptotically-flat solutions in the zero-temperature limit. The near-horizon metric includes an  AdS$_2$ factor with the extremal horizon located at $\rho=0$. As we discuss below, for fixed values of parameters $(J, Q, m, g)$, there may exist more than one distinct solution.

To understand how the NHG connects to the full extremal black hole, we study linear perturbations of Eq.~\eqref{eqs:nearh} generated by \emph{time-independent} deformations of the form $g \to g_{\rm NH}\left(1+ \rho^\gamma \,\delta g \right)$, $A \to A_{\rm NH}\left(1 + \rho^\gamma \,\delta A\right)$, and $a \to a_{\rm NH}\left(1 + \rho^\gamma \,\delta a\right)$, where all $\delta$ quantities are first order and depend only on the polar angle $\theta$. This ansatz admits a simple interpretation. Just as perturbations on a sphere can be decomposed into spherical harmonics, perturbations of AdS$_2$ may be expanded in harmonics on AdS$_2$. For time-independent modes, these harmonics reduce to power-law scalings $\propto \rho^\gamma$. Substituting this ansatz into the Einstein-Maxwell-axion equations yields a Sturm-Liouville-type problem whose spectrum determines the allowed values of $\gamma$ at fixed $Q$ and $J$.

There are three distinct possibilities. For each propagating degree of freedom: (i) there exists one positive and one negative value of $\gamma$, (ii) both values of $\gamma$ are negative, or (iii) $\gamma$ forms a complex conjugate pair~\footnote{Since the ${\rm AdS}_2$ Laplacian goes like $\partial_\rho (\rho^2 \partial_\rho)$ along the radial directions, the linearized mode equation depends on $\gamma$ only through the quadratic combination
$\lambda_\gamma\equiv\gamma(\gamma+1)$, so that determining $\lambda_\gamma$ fixes two values of $\gamma$. The problem can be recast as a Sturm-Liouville eigenvalue problem for $\lambda_\gamma$. {We can then show, using numerical methods, that the problem is self-adjoint and therefore possesses a real spectrum.} Consequently, the two corresponding values of $\gamma$ are either both real or form a complex conjugate pair, $\gamma=(-1\pm\sqrt{1+4\lambda_\gamma})/2$, which shows that when $\gamma$ is real the two roots necessarily lie on opposite sides of $-1/2$, forbidding the possibility that both are positive. The quantity $\lambda_\gamma$ plays a role analogous to the $SL(2,\mathbb{R})$ Casimir for fields on AdS$_2$, much as $\ell(\ell+1)$ does for spherical harmonics.}. The last case can be shown to lead to an instability of the full extremal solution~\cite{Durkee:2010ea,Hollands:2014lra}, while case~(ii) indicates that the NHG given in Eq.~\eqref{eqs:nearh} does not correctly describe the zero-temperature solutions. At first sight, option~(i) might appear to pose no difficulty, since one can impose boundary conditions that remove the negative $\gamma$. 
However, it was recently shown that if   $\gamma<2$ and $\gamma \ne 1$, the full spacetime geometry generically develops an infinite tidal-force singularity~\cite{Horowitz:2022mly,Horowitz:2022leb,Horowitz:2023xyl,Horowitz:2024dch,Horowitz:2024kcx,Horowitz:2025ayc}. Such tidal forces can be diagnosed from the components of the Riemann tensor in a frame associated with infalling timelike or null geodesics. 
As we discuss below, the main result of this paper is that for an axion, option (i) holds: $\gamma < 2$ and  $\gamma \ne 1$ for almost all axion parameters, so the associated extremal solutions are generically singular.

\medskip 

\begin{figure}
    \includegraphics[height=0.82\linewidth]{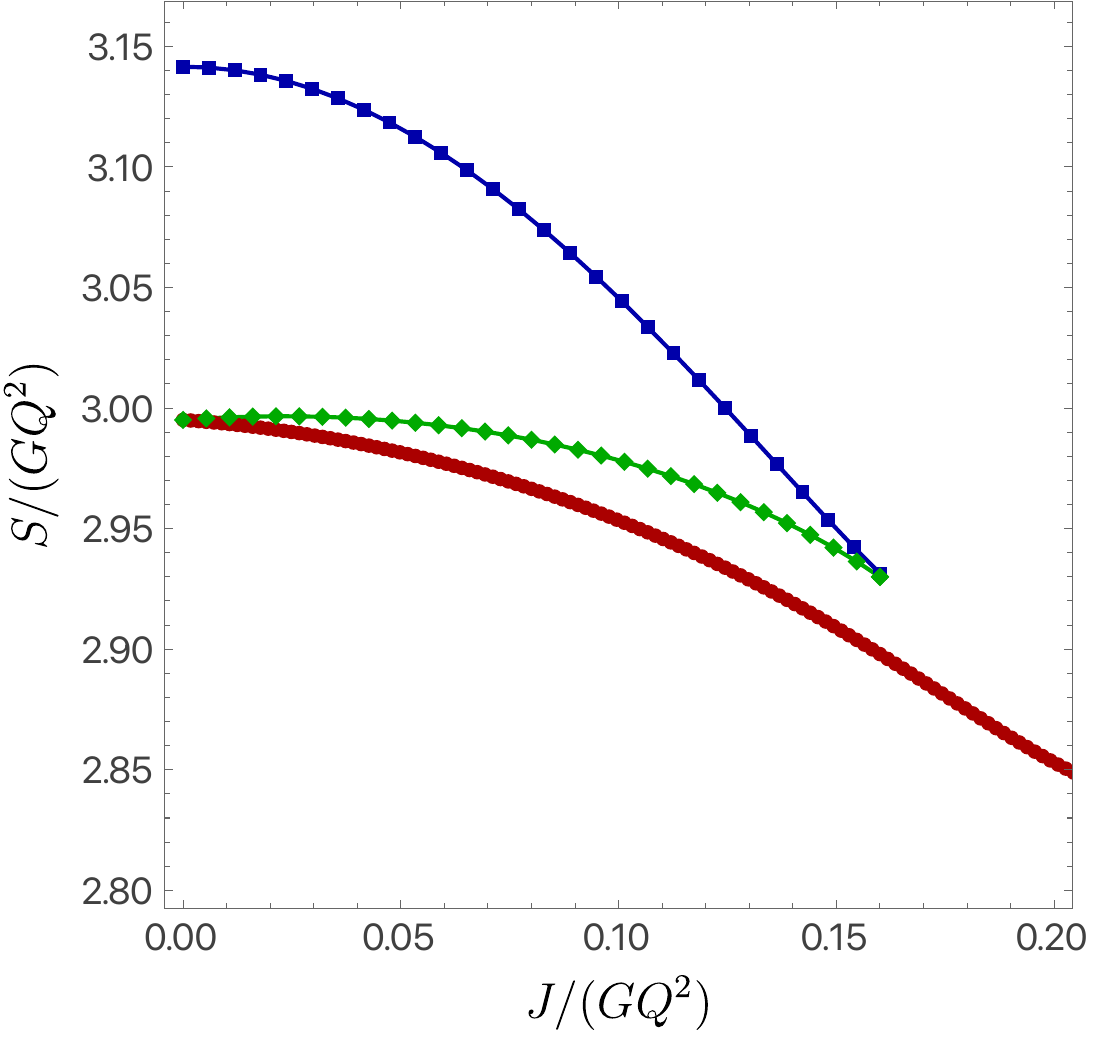}
    \caption{\label{fig:non}$S/(G Q^2)$ vs $ J/(GQ^2)$ for the near-horizon geometries with $g=10$ and $m G Q =7.5$. For small $ J/(GQ^2)$, three distinct branches are present. The upper branch connects to extreme Reissner-Nordstr\"om.}
\end{figure}

\mysec{Results}Our first result concerns the non-uniqueness of NHGs for our extremal black holes. In pure Einstein-Maxwell theory, the NHG is known to be unique~\cite{Lewandowski:2002ua}. With the axion, we find non-uniqueness. To the best of our knowledge, this is the first explicit example of such non-uniqueness in a four-dimensional theory with vanishing cosmological constant.  (Earlier work with massless axions had noticed non-uniqueness of the full non-extremal  solutions~\cite{Boskovic:2018lkj,Burrage:2023zvk}.) The non-uniqueness arises within a regime of parameter space characterized by small $ J/(GQ^2) $, sufficiently large $ g $, and sufficiently small $mGQ$. In particular, a perturbative analysis around the extremal Reissner-Nordstr\"om solution---using a spherical harmonic mode with angular momentum $\ell$---shows that non-uniqueness occurs when $mGQ < \sqrt{g^2 - \ell(\ell-1)}$, with $ \ell \geq 4 $~\footnote{In principle, $ \ell = 3 $ is also allowed; however, these modes break the discrete symmetry $ \theta \to \pi - \theta $ and will not be considered here.}. In Fig.~\ref{fig:non}, we plot the normalized entropy $S/(G Q^2)$ of three distinct NHGs with $g=10$ and $m G Q=7.5$, demonstrating that for $J/(GQ^2)\lesssim0.16$ there exist three distinct solutions. The upper branch (blue squares) connects to the extremal Reissner-Nordstr\"om black hole, while the two lower branches (green diamonds and red disks) approach a non-spherical NHG in the limit $J \to 0$. Among these, the lowest branch (red disks) ultimately connects to the NHG of an extremal Kerr black hole as $ J/(GQ^2) \to +\infty $. This branch is of primary astrophysical interest, since we expect $ GQ^2/J \ll 1 $.

\begin{figure}
    \includegraphics[height=0.82\linewidth]{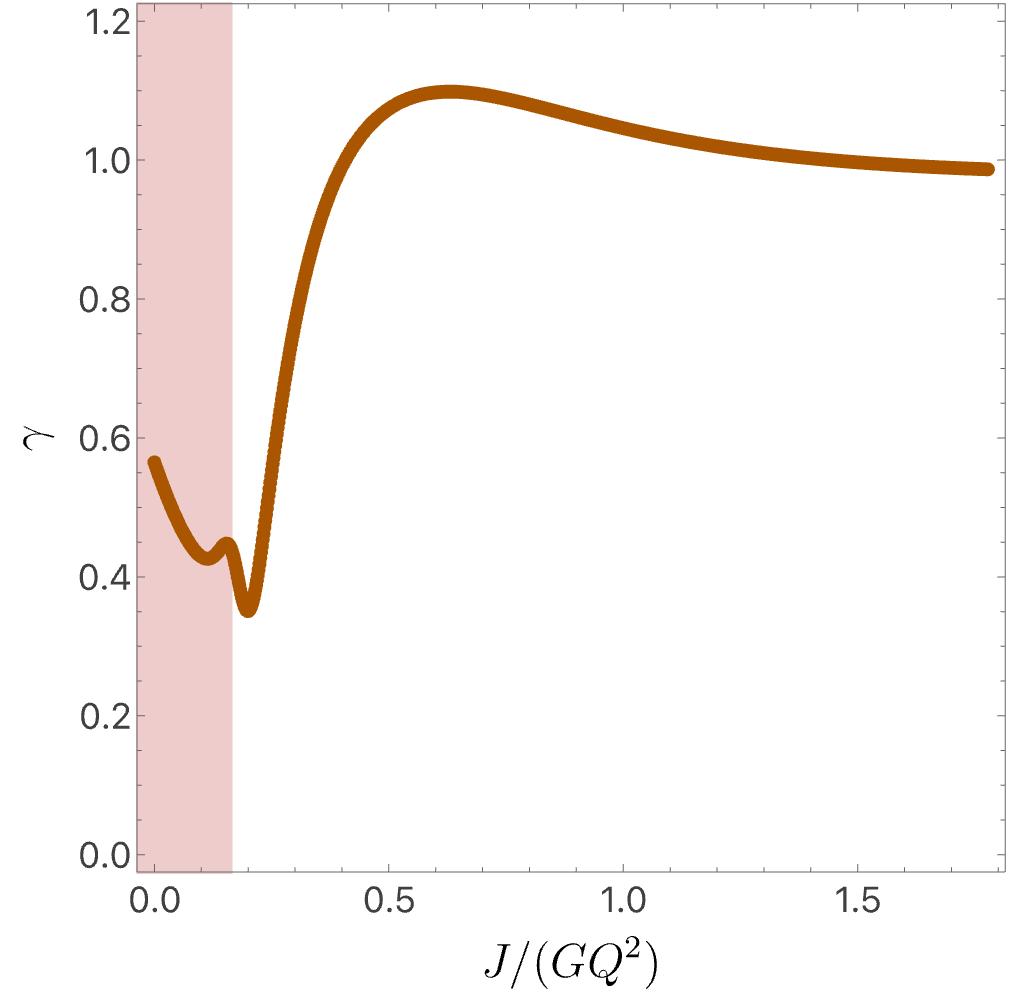}
    \caption{\label{fig:scaling}The lowest scaling dimension $\gamma$ vs. $J/(GQ^2)$ for $mGQ=7.5$, $g=10$. The shaded region indicates the regime of threefold degeneracy shown in Fig.~\ref{fig:non}.}
\end{figure}

We now turn to the scaling dimensions, which can be evaluated for each of the NHGs identified above. Focusing on the small-charge regime, we restrict attention to the family of solutions that extends to large $ J/(GQ^2)$, namely those connected in the $ (mGQ, J/(GQ^2), g) $ parameter space to the lower branch in Fig.~\ref{fig:non}. In Fig.~\ref{fig:scaling}, we show the lowest scaling dimension for $ mGQ = 7.5 $ and $ g = 10 $ as a function of $ J/(GQ^2) $. The red shaded region indicates the parameter range over which the threefold degeneracy identified in Fig.~\ref{fig:non} occurs. Over a wide region of parameter space, we find that $ \gamma < \gamma_0 $ where $\gamma_0 \approx 1.587513$. 
Indeed, this inequality holds throughout the $10^6$ points in $ (mGQ, J/(GQ^2), g) $ that we have sampled, with $ (g, mGQ) \in (0,10]^2$ and $|J|/(GQ^2)\in[0.2,10]$.  Note that in order to generate infinite tidal forces across the extremal horizon, we only need $\gamma<2$, which is clearly the case.

We have also verified that the mode identified in Fig.~\ref{fig:scaling} approaches $\gamma=1$ as $m G Q$ is increased at fixed $g$ and $J/(G Q^2)$. In fact,  $1-\gamma =O(1/m^2)$ (see Sec.~\hyperref[sup:5]{D} of the Supplemental Material), in agreement with the EFT analysis \cite{Horowitz:2024dch}. 

\begin{figure}
\includegraphics[height=0.82\linewidth]{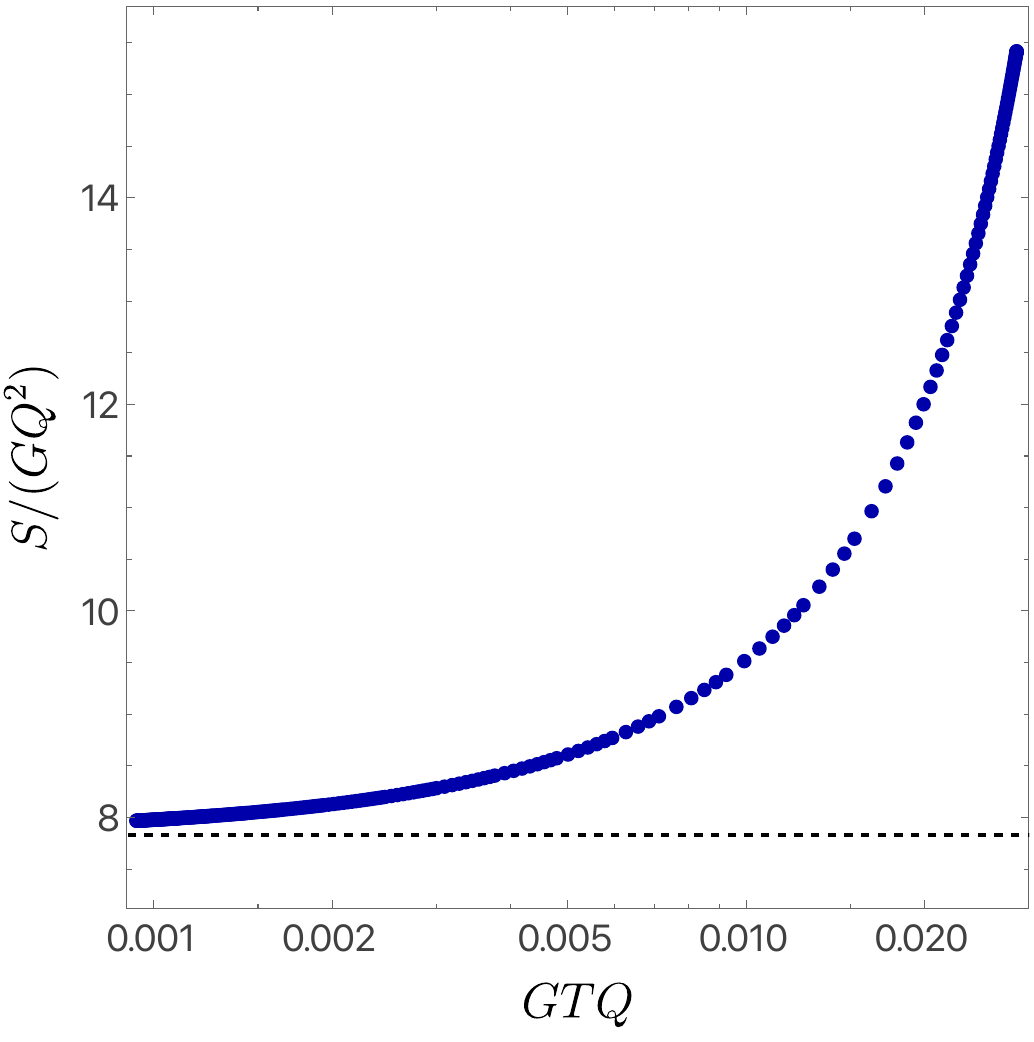}
    \caption{\label{fig:entropy}$S/(GQ^2)$ vs $TGQ$ for $mMG=10$, $Q/M=0.75$, and $g=10$. The dashed line denotes the near-horizon result, while the points correspond to the full finite-temperature solutions, showing agreement at low temperature.}
\end{figure}

Having established the existence of modes with $\gamma < 2$, one may ask whether these are necessarily excited in the full asymptotically-flat black hole geometry. We address this question in two steps. First, we find that the horizons of our finite-temperature solutions approach the NHGs identified in the zero-temperature limit. The zero-temperature limit is reached by holding $(mMG, Q/M, g)$ fixed while increasing $J$ to its maximum value $J_{\max}$. In Fig.~\ref{fig:entropy}, we plot the normalized entropy $S/(GQ^2)$, computed for $mMG = 10$, $Q/M = 0.75$, and $g = 10$, as a function of $GTQ$ on a linear-log scale. For these values we find that $J_{\max}/(G Q^2)\approx 1.18865$. The horizontal dashed line corresponds to the near-horizon result, while the blue points represent the finite-temperature computation. The agreement at low temperatures is reassuring. In Sec.~\hyperref[sup:4]{C} of the Supplemental Material, we further show that the Kretschmann scalar evaluated on the horizon matches the near-horizon prediction as the system is cooled.

Next, we compute the tidal forces across the future event horizon $\mathcal{H}^+$ along the equator $\theta = \pi/2$ at finite temperature and track their behavior as the system is cooled. These are defined by
\begin{equation}\label{eq:tidal}
\Phi \equiv \left.U^{\mu} U^\rho \xi^\nu \xi^\sigma R_{\mu \nu \rho \sigma}\right|_{\mathcal{H}^+}^{\theta=\pi/2}  ,
\end{equation}
where $U^a$ is tangent to ingoing affine null geodesics, normalized such that $U \cdot k = -1$ at large $r$. 

\begin{figure}
\includegraphics[height=0.82\linewidth]{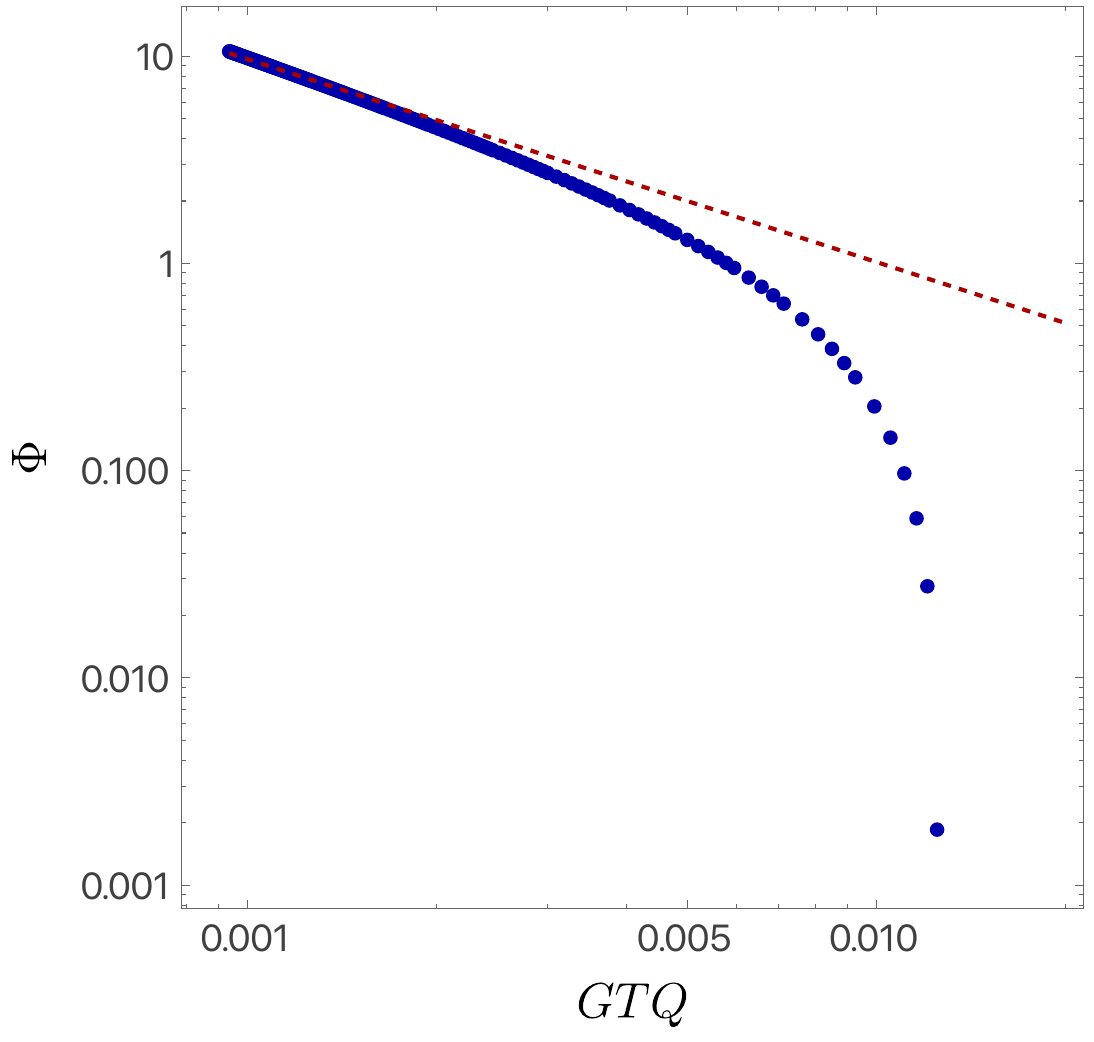}
    \caption{\label{fig:tidal} Log-log plot of the tidal force $\Phi$ in Eq.~\eqref{eq:tidal} vs. $GTQ$, showing the divergence as $T\to0$. The dashed line is a fit using the lowest scaling dimension from the near-horizon geometry. This is for $mMG=10$, $Q/M=0.75$, and $g=10$.}
\end{figure}

If our near-horizon analysis is correct, $\Phi$ should diverge as $T \to 0$ with a characteristic $1/T^{2-\gamma}$ scaling.  In Fig.~\ref{fig:tidal}, we plot $\Phi$ as a function of $GTQ$ on a $\log$-$\log$ scale, using the same parameters as in Fig.~\ref{fig:entropy}. For the corresponding value of $J_{\max}/(GQ^2)$ at $mGQ = 7.5$ and $g = 10$, the lowest scaling dimension extracted from Fig.~\ref{fig:scaling} is $\gamma \approx 1.021$. 
The red dashed line in Fig.~\ref{fig:tidal} shows a fit of the form $\alpha (GTQ)^{\gamma - 2}$ for $GTQ < 2 \times 10^{-3}$, with $\alpha \approx 0.011182$.

We have set $Q/M = 0.75$, but astrophysical black holes carry very small charge. For numerical purposes, the smallest charge we can reliably resolve while still accessing large spins is $Q/M = 0.01$. We again fix $g = 10$ and $mGM = 10$, which yields a near-horizon scaling exponent of $\gamma\approx 0.99997$. To quantify the size of tidal deformations, we compare our results with those of a Kerr-Newman black hole with the same values of $Q/M$ and $J/(GM^2)$, denoting the corresponding quantity by $\Phi_{\rm K}$. For the parameters above, we find that $(\Phi - \Phi_{\rm K})/\Phi_{\rm K} \approx 5 \times 10^{-8}$ at the Thorne limit~\cite{Thorne} of $J/(GM^2) \approx 0.998$. 

\begin{figure}
    \includegraphics[height=0.82\linewidth]{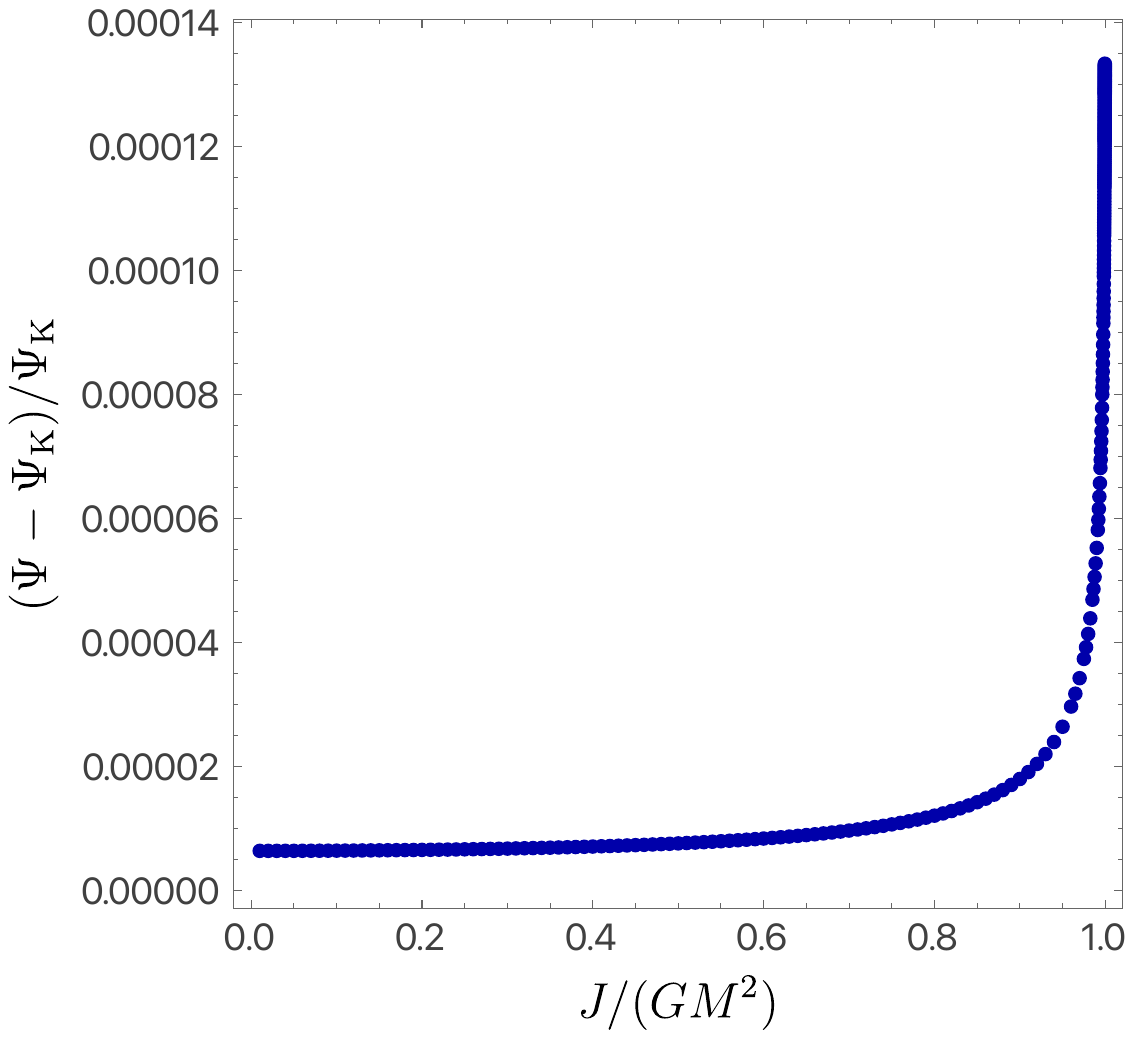}
    \caption{\label{fig:tidal_maxwell}
    Relative deviation of the electric field in Eq.~\eqref{eq:Efield} from Kerr-Newman as a function of $J/(GM^2)$, plotted for $Q/M = 0.01$, $g = 10$, and $mGM = 10$.}
\end{figure}

Since $\gamma < 1$, we also find a divergence in the electric field seen by an infalling observer. We set
\begin{equation}\label{eq:Efield}
\Psi \equiv \left.U^\mu \xi^\nu F_{\mu \nu}\right|_{\mathcal{H}^+}^{\theta=\pi/2}
\end{equation}
and compare it to the Kerr-Newman result $\Psi_{\rm K}$. In this case, we find $(\Psi - \Psi_{\rm K})/\Psi_{\rm K} \approx 10^{-4}$ at $J/(GM^2) \approx 0.998$. We plot this quantity in Fig.~\ref{fig:tidal_maxwell}, where the rapid growth near extremality is visible. 

\medskip

\mysec{Discussion}We have shown that an  axion---one of the leading dark-matter candidates---causes rotating charged black holes to have singular horizons in their extremal limit. Both rotation and charge are important for this effect. Astrophysical black holes can have a small nonzero charge induced by an external magnetic field~\cite{Wald:1974np}, so
this result opens the door to placing new and potentially stringent constraints on the axion coupling $g$ and mass $m$. However, our current estimates indicate that to reach anomalously large tidal forces one must be much closer to extremality than is 
expected  for an astrophysical black hole. 

While we have focused on the case of massive axions, similar results should hold for massive dilatons. In other words, even if a dilaton is quite massive, extremal rotating charged black holes should be expected to have singular horizons. (Earlier discussions of black holes with a massive dilaton~\cite{Horne:1992bi,Gregory:1992kr} did not see this effect, since they only considered non-rotating black holes.)
Indeed, in the large-mass limit, the axion and the dilaton may be integrated out, leaving an Einstein-Maxwell effective theory with $F^4$ terms, which generate these singularities as we found in Ref.~\cite{Horowitz:2024dch}.

This work leaves various compelling directions for further investigation.
One open question is whether there are situations in which the anomalously large tidal forces are enhanced so that they might be present for astrophysical black holes. On a more theoretical level, another open question is to understand how the quantum Schwarzian modes (which become important very close to extremality) affect our classical tidal force singularity.

\bigskip
\noindent {\it Acknowledgments:} 
G.~H. and M.~K. were supported in part by NSF grant PHY-2408110. G.~H. was also supported in part by Simons Foundation International and the Simons Foundation through Simons Foundation grant SFI-MPS-BH-00012593-08. J.~E.~S. is partially supported by STFC grant ST/X000664/1 and by Hughes Hall College. G.~N.~R. is supported by the James Arthur Postdoctoral Fellowship at New York University.

\bibliographystyle{utphys-modified}
\bibliography{axionBH}

\onecolumngrid
\section*{Supplemental Material}

\subsection{\label{sup:2}Conformal coordinates}
We choose a gauge where
\begin{equation}
Q_2(r,\theta)=\frac{f(r)}{B(r,\theta)},
\qquad
Q_8(r,\theta)=B(r,\theta),
\end{equation}
with
\begin{equation}
f(r)=\left(1-\frac{r_+}{r}\right)\left(1-\frac{\bar{Q}^2}{r\,r_+}\right).
\end{equation}
In this parameterization, the event horizon is located at $r=r_+$, while the constant $\bar{Q}$ is a gauge parameter to be discussed below.

Furthermore, we perform a redefinition of the functions $Q_i$ such that the metric, gauge field, and axion ansatz take the form
\begin{equation}
{\rm d}s^2 = -f(r) A(r,x){\rm d}t^2
+ (1-x^2)r^2S(r,x)\left[{\rm d}\phi-\frac{r_+^3}{r^3}W(r,x){\rm d}t\right]^2
+ B(r,x)\,\left[\frac{{\rm d}r^2}{f(r)} + \frac{r^2{\rm d}x^2}{1-x^2}\right],
\end{equation}
\begin{equation}
A=A_t(r,x)\left(1-\frac{r_+}{r}\right){\rm d}t+A_\phi(r,x)(1-x^2)\left[{\rm d}\phi-\frac{r_+^3}{r^3}W(r,x){\rm d}t\right],
\end{equation}
and
\begin{equation}
a=\varphi(r,x),
\end{equation}
where we have introduced the coordinate $x \equiv \cos\theta \in [-1,1]$. In the special case where $A=B=S=1$ and $W=0$, the above line element describes a Reissner-Nordstr\"om black hole of mass $M=(r_+^2+\bar{Q}^2)/(2 G r_+)$ and electric charge $Q=\bar{Q}/G$, which will be a solution so long as $g=0$ and $A_t=\bar{Q}/r_+$.

\subsection{\label{sup:3}Numerical methods}
Our numerical method closely follows Sec.~5 of Ref.~\cite{Horowitz:2024kcx}. In particular, we solve the following components of the Einstein equation in Eq.~\eqref{eq:eomein},
\begin{equation}
{\bf E}\equiv\{E_{tt},E_{t\phi},E_{\phi\phi},g^{rr}E_{rr}+g^{xx}E_{xx}\},
\end{equation}
together with the equations of motion $E_{\mu}=0$ for the gauge field in Eq.~\eqref{eq:eomA} along with the equation for the axion in Eq.~\eqref{eq:eomphi}. This yields a total of seven partial differential equations in $(r,x)$ for the functions $\{A,B,S,W,A_t,A_{\phi},\varphi\}$. We refer to this restricted set of equations as the \emph{dynamical equations}.

The remaining nonvanishing components of the Einstein equations, $C^1 \equiv E_{rx}=0$ and $C^2 \equiv g^{rr}E_{rr}-g^{xx}E_{xx}=0$, can be shown to satisfy a Cauchy-Riemann-type system once the dynamical equations are imposed. We therefore treat $\{C^1,C^2\}$ as constraint equations and monitor them as consistency checks of the numerical solution.

It is convenient to work with a compact coordinate,
\begin{equation}
r=\frac{r_+}{1-y},
\end{equation}
so that asymptotic infinity is located at $y=1$ and the horizon at $y=0$. In Ref.~\cite{Horowitz:2024kcx}, appropriate boundary conditions at $y=0$ and $x=\pm 1$ were identified that define a well-posed elliptic problem with locally unique solutions.

For our purposes, it is convenient to work at fixed total charges $\{M,Q,J\}$. To the best of our knowledge, this choice has not been previously used to numerically determine solutions to the Einstein equations at the nonlinear level. To identify the energy, we compute the Komar integral,
\begin{equation}
M= -\lim_{r\to+\infty}\frac{1}{\kappa^2}\int_{S^2_{tr}}
\left\{\star\!\left[\mathrm{d}k^\flat + 2\,(\iota_k A) \bar{F}\right]- 2\,(\iota_k \bar{F})\wedge A
\right\},
\end{equation}
with $k=\partial/\partial t$ and $k^{\flat}\equiv k_\mu {\rm d}x^{\mu}$. Unlike the angular momentum and electric charge, $M$ is \emph{not} conserved on $S^{2}_{tr}$ hypersurfaces due to a ``bulk'' term proportional to $m^2$.

The boundary conditions at spatial infinity, located at $y=1$, read as follows:
\begin{equation}
\begin{aligned}
q_1(x,1)&=1\\
\left.\frac{\partial^2q_2}{\partial y^2}\right|_{y=1}+(1-2 x^2) \left.\frac{\partial q_3}{\partial y}\right|_{y=1}-\frac{1}{4} q_3(x,1) (1-x^2) \left.\frac{\partial^2q_2}{\partial y^2}\right|_{y=1}\hspace{44mm}
\\
+\frac{5}{2} (1-x^2) \left[\hat{Q}^2+q_3(x,1) \left(1+\left.\frac{\partial q_1}{\partial y}\right|_{y=1}\right)-\hat{M}
   \sqrt{q_3(x,1)} \left(2+\left.\frac{\partial q_1}{\partial y}\right|_{y=1}\right)\right]&=0\\
\left.\frac{\partial q_1}{\partial y}\right|_{y=1}+1-\frac{2 \hat{M}}{\sqrt{q_3(x,1)}}+\tilde{Q}^2&=0\\
q_4(x,1)-\frac{2 \hat{J}}{q_3(x,1)^{3/2}}&=0\\
q_5(x,1)&=0\\
\left.\frac{\partial q_6}{\partial y}\right|_{y=1}+q_6(x,1)-\frac{\hat{Q}}{\sqrt{q_3(x,1)}}&=0\\
q_7(x,1)&=0,
\end{aligned}
\end{equation}
where we introduced $\hat{Q} \equiv G Q$, $\hat{M}\equiv G M$, $\hat{J}\equiv G J$, $\tilde{Q}\equiv \hat{Q}/r_+$, and
\begin{equation}
\begin{aligned}
q_1(x,y)&\equiv A\left(\frac{r_+}{1-y},x\right),&\quad q_2(x,y)&\equiv B\left(\frac{r_+}{1-y},x\right),&\quad q_3(x,y)&\equiv S\left(\frac{r_+}{1-y},x\right),
\\
q_4(x,y)&\equiv W\left(\frac{r_+}{1-y},x\right), &\quad q_5(x,y)&\equiv \varphi\left(\frac{r_+}{1-y},x\right),&\quad q_6(x,y)&\equiv A_{t}\left(\frac{r_+}{1-y},x\right),
\\
q_7(x,y)&\equiv A_{\phi}\left(\frac{r_+}{1-y},x\right).
\end{aligned}
\end{equation}
Our boundary conditions require $q_3(x,1)=q_2(x,1)=\eta$, where $\eta$ is a constant. This follows from the absence of conical singularities at $x=\pm 1$, together with an expansion near $y=1$. However, we do not fix $\eta=1$. This is because we work at fixed mass $M$, a choice more conveniently implemented by allowing $\eta$ to vary at $y=1$. Nonetheless, the condition $q_3(x,1)=q_2(x,1)=\eta$ still guarantees asymptotic flatness at spatial infinity.
The boundary conditions at the remaining edges of the integration domain are determined by regularity at $y=0$ (the horizon) and by the absence of conical singularities along the axis $x=\pm 1$~\cite{Dias:2015nua}.

To solve the dynamical equations, we discretize the system using Gauss-Lobatto-Chebyshev grids in both directions, with $x \in [-1,1]$ and $y \in [0,1]$, employing $N_x$ and $N_y$ collocation points, respectively. This spectral collocation method provides high accuracy for smooth solutions and is well suited to the present problem. The resulting set of nonlinear algebraic equations is then solved using a standard Newton-Raphson iteration scheme implemented in extended precision, which ensures numerical stability and convergence. Further details on these numerical techniques and their implementation can be found in Ref.~\cite{Dias:2015nua}.

\begin{figure}
    \includegraphics[width=0.69\linewidth]{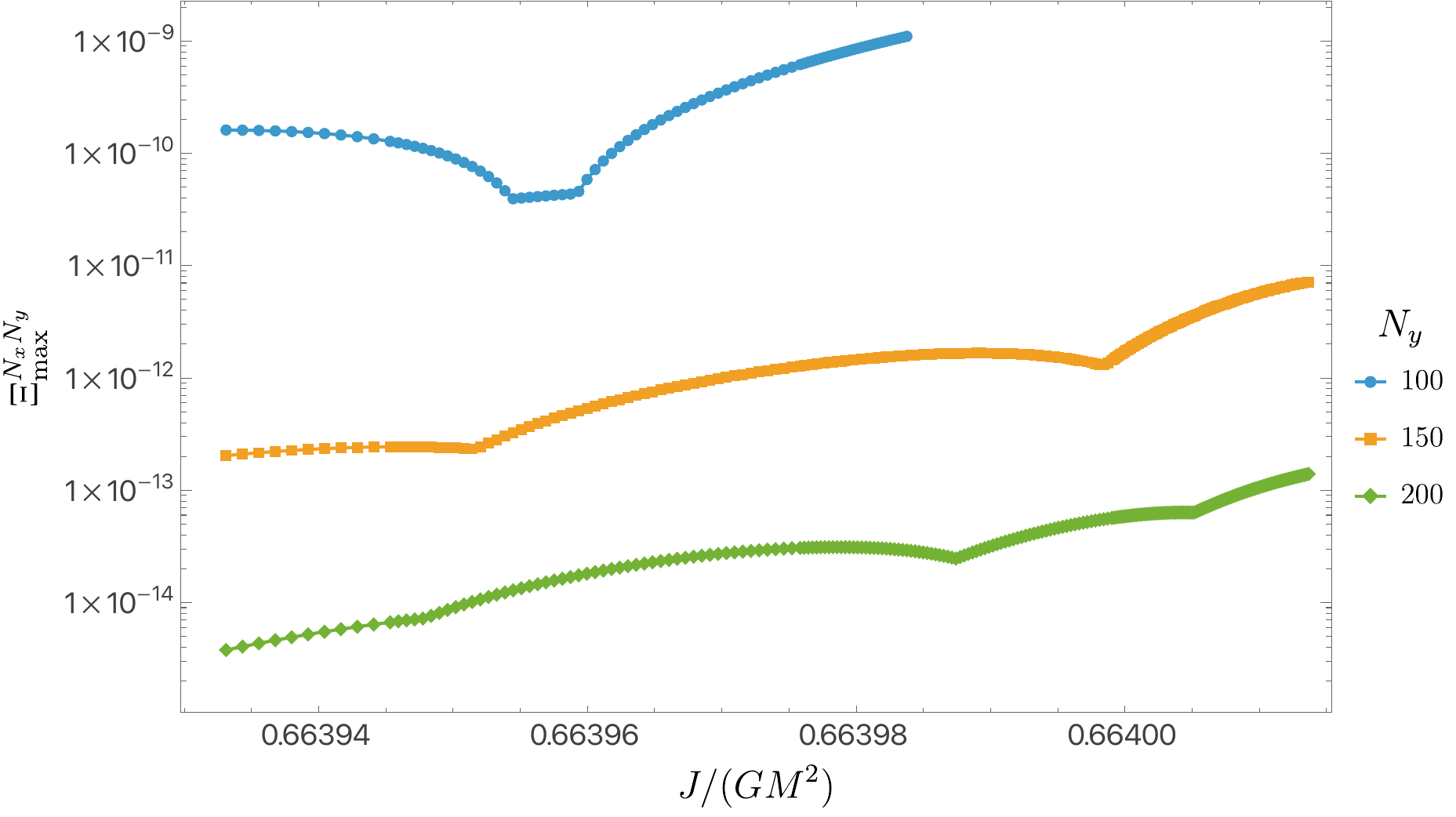}
    \caption{\label{fig:example_con}Convergence test quantity $\Xi_{\max}^{N_x N_y}$ as a function of $J/(G M^2)$, computed at several resolutions $N_y$  (indicated on the right). The curves shown correspond to $Q/M = 0.75$, $mGM = 1$, $g = 1$, and $N_x = 30$.}
\end{figure}

As a check of our numerical procedure, we discretize the constraints ${C^1, C^2}$ using the same Gauss-Lobatto-Chebyshev grids employed for solving the dynamical equations. This yields the discrete constraint matrices $C^{1,N_x N_y}_{ij}$ and $C^{2,N_x N_y}_{ij}$, which we assemble into a single vector-valued object,
\begin{equation}
\boldsymbol{\Xi}^{N_x N_y}
\equiv\bigl\{C^{1,N_x N_y}_{ij}\,, C^{2,N_x N_y}_{ij} \bigr\},
\end{equation}
where $i$ and $j$ index the discretization points. We then define
\begin{equation}
\Xi_{\max}^{N_x N_y}\equiv \lVert \boldsymbol{\Xi}^{N_x N_y} \rVert_{\infty},
\end{equation}
and study the behavior of $\Xi_{\max}^{N_x N_y}$ as the resolution is increased, i.e., as a function of $N_x$ and $N_y$. In all cases, $\Xi_{\max}^{N_x N_y}$ exhibits exponential convergence with increasing resolution $(N_x, N_y)$, as expected for spectral collocation methods on Gauss-Lobatto-Chebyshev grids applied to smooth functions. See Fig.~\ref{fig:example_con} for an example of a convergence test focusing near the region close to extremality. 

\subsection{\label{sup:4}Further evidence for matching} 
In the main text, we provide evidence for the matching between the entropy computed directly in the NHG at zero temperature, at fixed $m G Q$, $J/(G Q^2)$, and $g$, and the low-temperature limit of the entropy of the full spacetime at finite temperature, $G T Q > 0$. One may then ask whether other local quantities characterizing the geometry also exhibit similar agreement.

\begin{figure}
    \includegraphics[width=0.6\linewidth]{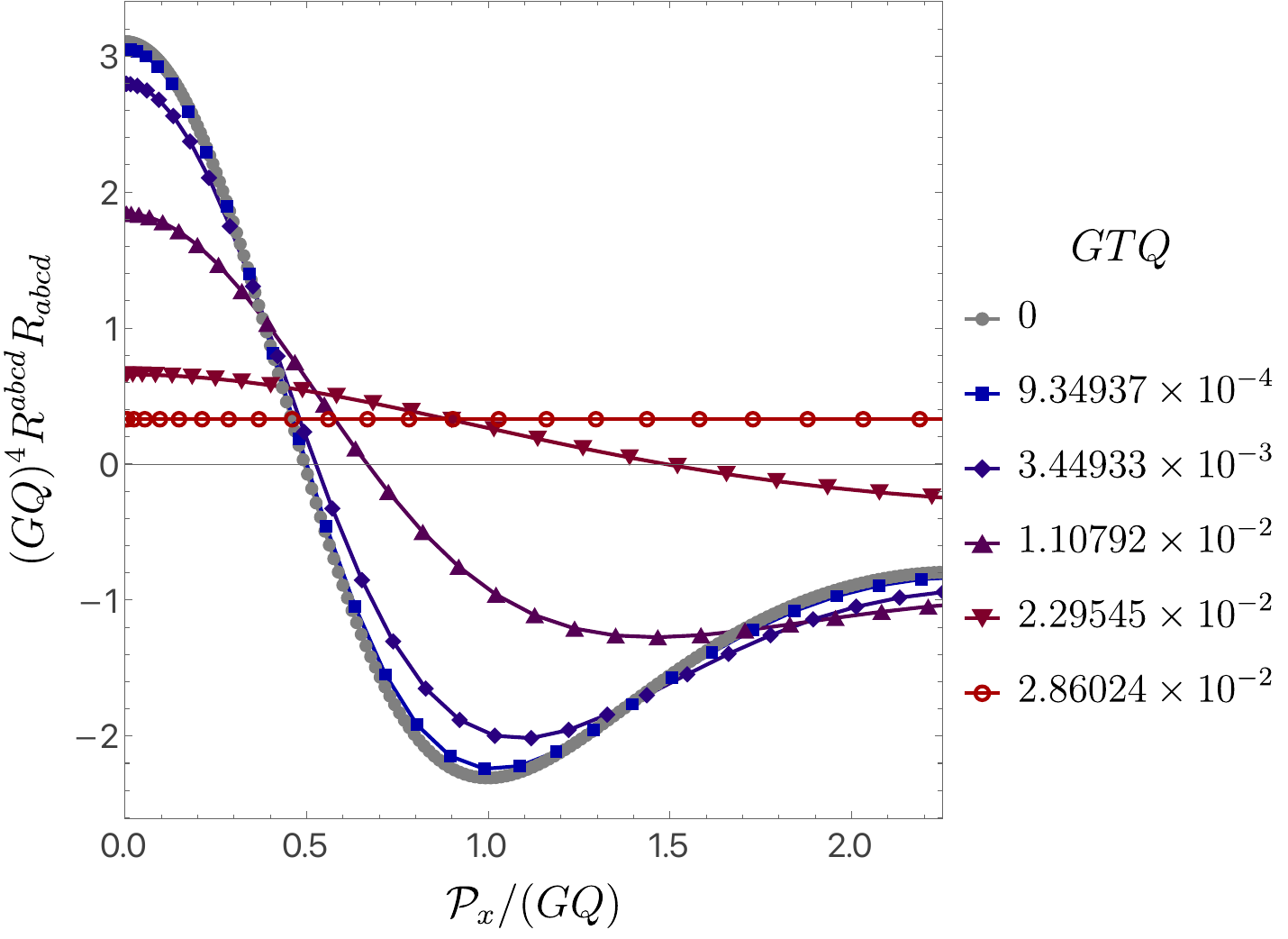}
    \caption{\label{fig:curvature_approach}Kretschmann scalar evaluated on the black hole horizon as a function of proper distance from the equator ($x=0$). Results at finite temperature (labeled on the right) are compared with the zero-temperature solution, showing excellent agreement. The finite-temperature solutions are computed at fixed $mGM = 10$, $Q/M = 0.75$, and $g = 10$. For the zero-temperature solution, we use $J/(G M^2) \approx 1.18865$.}
\end{figure}

In Fig.~\ref{fig:curvature_approach}, we plot the spacetime Kretschmann scalar $R_{\mu\nu\rho\sigma}R^{\mu\nu\rho\sigma}$ evaluated on the black hole horizon as a function of the proper distance $\mathcal{P}_x$ along the horizon from the equator, $\theta = \pi/2$ (corresponding to $x = 0$), i.e.,
\begin{equation}
\mathcal{P}_x=\int_0^x{\rm d}\tilde{x}\,\sqrt{\frac{q_2(\tilde x,0)}{1-\tilde x^2}}.
\end{equation}
The zero-temperature solution is obtained using the largest value of $J/(G M^2) \approx 1.18865$ reached in the finite-temperature construction. The finite-temperature solutions are computed at fixed $mGM = 10$, $Q/M = 0.75$, and $g = 10$. The agreement is excellent.

\subsection{\label{sup:5}Matching with EFT }
\label{sup:5}
At large $m G Q$, while holding $J/(G Q^2)$ and $g$ fixed, we expect to recover the EFT results of Ref.~\cite{Horowitz:2024dch}. In that work, two distinct classes of near-horizon modes were identified prior to adding EFT corrections: those with $\lambda^{(0)} = 2$ (equivalently $\gamma^{(0)} = 1$), and those with $\lambda^{(0)} > 2$ (i.e., $\gamma^{(0)} > 1$). This distinction arises because modes with $\gamma^{(0)} = 1$ require a field ansatz for the EFT perturbations that includes terms of the form $\rho \log \rho$, in addition to the usual $\rho^{\gamma^{(0)}}$ behavior. Indeed, it was shown in Ref.~\cite{Horowitz:2024dch} that such $\rho \log \rho$ terms are unavoidable, even for the Kerr-Newman solution, when expressed in the coordinate system adopted there.

\begin{figure}
    \includegraphics[width=0.9\linewidth]{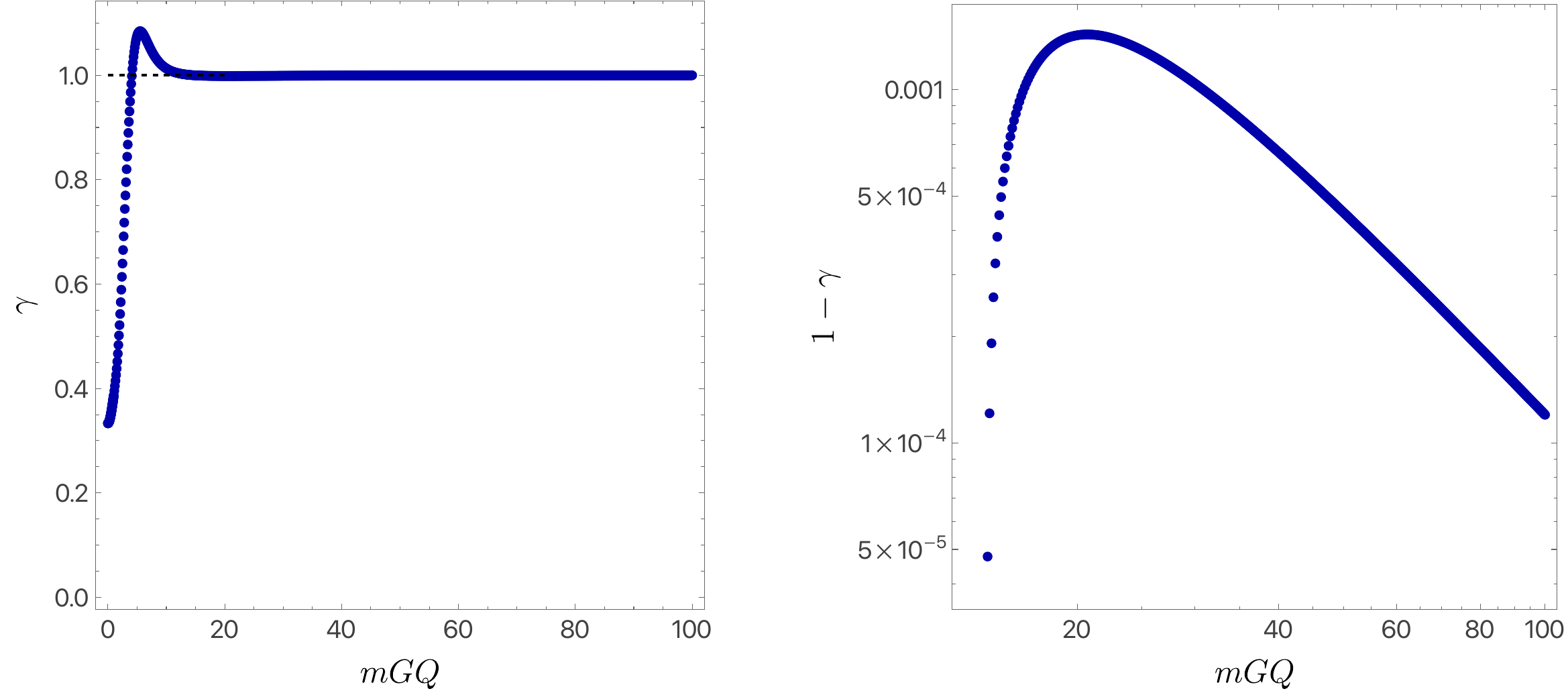}
    \caption{\label{fig:gamma1}Left: $\gamma$ as a function of $m G Q$ for fixed $g=10$ and $J/(G Q^2)=1$. Right: $1-\gamma$ as a function of $m G Q$ on a $\log$-$\log$ scale. The approach to $\gamma=1$ is consistent with a $(m G Q)^{-2}$ scaling, as expected in the EFT~\cite{Horowitz:2024dch}.}
\end{figure}

The EFT modes with $\gamma^{(0)} = 1$ give rise to a tidal force that diverges as $1/T$ within the EFT, but vanishes for both Kerr and Reissner-Nordstr\"om black holes. We therefore expect that, in the spectrum of axion black holes, one mode approaches $\gamma = 1$ in the limit of large $m G Q$, at fixed $J/(G Q^2)$ and $g$.
For large axion masses, we can integrate out the axion, yielding $F^4$ terms considered in Ref.~\cite{Horowitz:2024kcx}, with Wilson coefficients $\propto 1/m^2$.
The analysis of Ref.~\cite{Horowitz:2024dch} predicts that the $\gamma^{(0)}=1$ modes will have scaling dimension that approaches $1$ at a rate proportional to $(m G Q)^{-2}$. However, the multiplicative coefficient is not determined due to the presence of $\rho \log \rho$ terms in the near-horizon expansion, which obscure the identification of the quantity denoted $\delta\gamma$ in Ref.~\cite{Horowitz:2024dch}.
In the left panel of Fig.~\ref{fig:gamma1}, we plot $\gamma$ as a function of $m G Q$ for fixed $g=10$ and $J/(G Q^2)=1$. In the right panel, we show $1-\gamma$ as a function of $m G Q$ on a $\log$-$\log$ scale. The approach to $\gamma=1$ is fully consistent with a $(m G Q)^{-2}$ scaling, as expected from the EFT~\cite{Horowitz:2024dch}.

\begin{figure}
    \includegraphics[height=0.41\columnwidth]{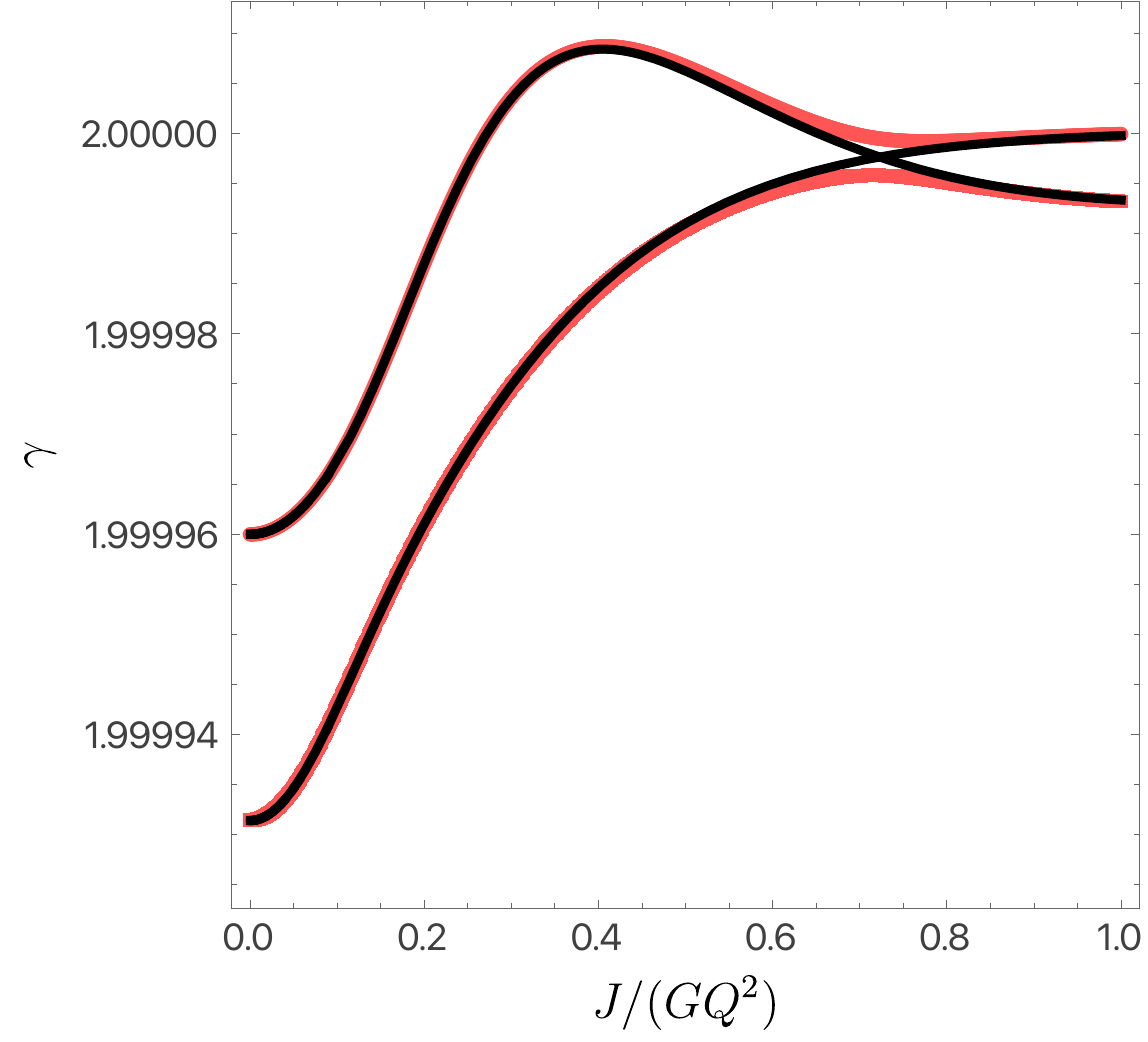}
    \caption{\label{fig:comp1} Scaling dimension $\gamma$ of the second-lowest mode versus $J/(G Q^2)$ for $g=1$ and $mGQ=100$. The solid black curves correspond to the analytic EFT corrections $\delta\gamma^{(K)}_{\pm}$ about $\gamma^{(0)}\,{=}\,2$ predicted by Ref.~\cite{Horowitz:2024dch}, while light red disks show our numerical data. The agreement is excellent. The apparent crossing of the analytic branches is resolved through eigenvalue repulsion, consistent with the behavior observed in Ref.~\cite{Horowitz:2024kcx}.}
\end{figure}

For higher modes with $\gamma^{(0)} > 1$,  a detailed matching at large $m G Q$ can be carried out, as we now demonstrate. In order to provide a detailed matching, we need to first identify the Wilson coefficients of the EFT operators appearing in Ref.~\cite{Horowitz:2024dch} in terms of the axion coupling $g$ and $m$. Integrating out the axion at tree level, we find, in the terminology of Ref.~\cite{Horowitz:2024dch},
\begin{equation}
   (d_7,d_8) =\frac{1}{8\pi G} (c_7, c_8) =  \frac{g^2}{32m^2} \left(-1,2
    \right).
\end{equation}
and $d_6=0$. In Fig.~\ref{fig:comp1}, we plot $\gamma$ as a function of $J/(G Q^2)$ for $g=1$ and $mGQ=100$. According to Ref.~\cite{Horowitz:2024dch}, modes with $\gamma^{(0)}=2$ receive two corrections, $\delta \gamma^{(K)}_{\pm}$, which are shown in Fig.~\ref{fig:comp1} as solid black lines. The light red disks correspond to the values of $\gamma$ obtained from the axion-corrected NHG. The quantitative agreement is excellent. Notably, the corrections $\delta \gamma^{(K)}_{\pm}$ intersect, and this crossing is resolved through eigenvalue repulsion, similar to what was observed in Ref.~\cite{Horowitz:2024kcx}.

\subsection{\label{sup:6}Perturbative expansion in $G Q^2/J$ for the massless axion}
For small $G Q^2/J$ with $m=0$, one can employ perturbation theory to make analytic progress. The metric and gauge field can be determined up to order $G^2 Q^4/J^2$ for arbitrary temperature, charge, and axion coupling $g$. By contrast, the axion field can only be consistently determined to this order in the \emph{extremal} limit. In this case, we find that the axion is non-smooth across the future event horizon. We expect that this non-smoothness will backreact on the geometry, contributing to the metric at higher orders in the perturbative expansion.

In order to make progress, we take the following metric and gauge field ansatz:
\begin{equation}
\begin{aligned}
{\rm d}s^2=&\,-\frac{\Delta(r)}{\Sigma(r,x)}\tilde{A}(r,x)[{\rm d}t-(1-x^2)\tilde{W}(r,x) {\rm d}\phi]^2+\frac{1-x^2}{\Sigma(r,x)}\tilde{S}(r,x)[\tilde{W}(r,x){\rm d}t-(r^2+a^2){\rm d}\phi]^2
\\
&\, +\Sigma(r,x)\tilde{B}(r,x)\left[\frac{{\rm d}r^2}{\Delta(r)}+\frac{{\rm d}x^2}{1-x^2}\right]\\
A=&\,-\frac{\tilde{A}_t(r,x)}{\Sigma(r,x)}[{\rm d}t-(1-x^2)\tilde{W}(r,x) {\rm d}\phi]-\frac{\tilde{A}_{\phi}(r,x)}{\Sigma(r,x)}[\tilde{W}(r,x){\rm d}t-(r^2+a^2){\rm d}\phi]\\
a=&\,\tilde{\varphi}(r,x).
\end{aligned}
\end{equation}
A purely electrically charged Kerr-Newman black hole is obtained from the above by setting
\begin{equation}
\tilde{A}=\tilde{B}=\tilde{S}=1,\quad \tilde{W}=a,\quad \tilde{A}_t=\hat{Q} r,\quad \tilde{A}_{\phi}=0,
\end{equation}
and
\begin{equation}
\Delta(r)=r^2-2\tilde{M}r+a^2+\hat{Q}^2\,\quad \text{and}\quad \Sigma(r,x)=r^2+a^2x^2,
\label{eq:fix}
\end{equation}
where the Kerr-Newman black hole mass, electric charge, and angular momentum are given by $M=\tilde{M}/G$,  $Q=\hat{Q}/G$, and $J=aM$, respectively.

We then set
\begin{equation}
\begin{aligned}
\tilde{A}(r,x)&=1+\sum_{i=1}^{\infty}\hat{Q}^{2i}\tilde{A}^{(2i)}(r,x),&\;\; \tilde{B}(r,x)&=1+\sum_{i=1}^{\infty}\hat{Q}^{2i}\tilde{B}^{(2i)}(r,x),&\;\; \tilde{S}(r,x)&=1+\sum_{i=1}^{\infty}\hat{Q}^{2i}\tilde{S}^{(2i)}(r,x)
\\
\tilde{W}(r,x)&=a+\sum_{i=1}^{\infty}\hat{Q}^{2i}\tilde{W}^{(2i)}(r,x),&\;\; \tilde{A}_t(r,x)&=\hat{Q} r+\sum_{i=1}^{\infty}\hat{Q}^{2i+1}\tilde{A}_t^{(2i)}(r,x),&\;\; \tilde{A}_\phi(r,x)&=\sum_{i=1}^{\infty}\hat{Q}^{2i+1}\tilde{A}_{\phi}^{(2i)}(r,x),
\\
\tilde{\varphi}&=\sum_{i=1}^{\infty}\hat{Q}^{2i}\tilde{\varphi}^{(2i)}(r,x),
\end{aligned}
\end{equation}
and fix $\Delta$ and $\Sigma$ as in Eq.~(\ref{eq:fix}). Note that, as we work in perturbation theory, $\tilde{M}$ will no longer be directly proportional to the black hole mass. Instead, we should regard $\tilde{M}$ as parameterizing the horizon radius $r=r_+$ (to all orders in perturbation theory) as
\begin{equation}
\tilde{M}=\frac{r_+^2+a^2+\hat{Q}^2}{2 r_+}.
\end{equation}

In our perturbation scheme we demand that $a=J/M$, $\lim_{r\to +\infty} \tilde{A}^{(2i)}(r,x)=0$, and $Q=\hat{Q}/G$ to all orders in $Q$. Note that the absence of conical singularities at $x=\pm1$ also demands $B^{(2i)}(r,\pm1)=S^{(2i)}(r,\pm1)$.
To quadratic order in $\hat{Q}/\sqrt{J}$, we find that
\begin{equation}
A^{(2)}(r,x)=B^{(2)}(r,x)=S^{(2)}(r,x)=W^{(2)}(r,x)=0,
\end{equation}
which implies that the Kerr-Newman black hole is still a solution to this order in perturbation theory. The axion, however, becomes
\begin{equation}
\tilde{\varphi}^{(2)}(r,x)=\frac{a g x}{4 \tilde{M} \left(r^2+a^2 x^2\right)}.
\end{equation}

At third order in perturbation theory, the gauge field is modified to
\begin{equation}
\tilde{A}^{(3)}_t(r,x)=-\frac{a^2g^2(r+2\tilde{M} x^2-r x^2)}{24 \tilde{M}^2(r^2+a^2 x^2)}, \qquad \tilde{A}^{(3)}_{\phi}(r,x)=\frac{a g^2 r(1-x^2)}{24 \tilde{M}^2(r^2+a^2x^2)}.
\end{equation}

Finally, at fourth order in perturbation theory, both the metric and axion receive corrections. We were able to determine these corrections for the metric in closed form, which turn out to be complicated rational functions of $r$ and $x$. For completeness, these are as follows: 
\begin{equation}
\begin{aligned}
\tilde{A}^{(4)}(r,x)&=\frac{a^2 g^2}{48 \tilde{M}^6 r (r^2+a^2 x^2)^2}\Bigg\{\tilde{M} r^3 (r+\tilde{M})+2 \tilde{M}^4 r x^2+a^4 (r+\tilde{M}) x^2 (1-x^2)
\\
& \hspace{44mm} +a^2 \left[r-(r-\tilde{M}) x^2\right] \left[\tilde{M} r+r^2+2 \tilde{M}^2 (1-x^2)\right]\Bigg\}\\
\tilde{B}^{(4)}(r,x)&=-\frac{a^2 g^2}{48 \tilde{M}^5 (r^2+a^2 x^2)^2}\left[2 \tilde{M}^3 x^2+(r^2+a^2 x^2) (r+\tilde{M} x^2)\right]\\
\tilde{S}^{(4)}(r,x)&=\frac{a^2 g^2}{48 \tilde{M}^6 r (r^2+a^2 x^2)^2}\Bigg\{
a^4 (r+\tilde{M}) x^2 (1-x^2)-\tilde{M} r \left[r (4\tilde{M}^2+\tilde{M} r+r^2)+2 \tilde{M}^2 (\tilde{M}-2 r) x^2\right]
\\
& \hspace{31mm} +a^2 \left[r^3 (1-x^2)+2 \tilde{M}^3 x^2 (1-x^2)+\tilde{M} r^2 (1-2 x^2)+\tilde{M}^2 r(2-x^2) (1-2 x^2)\right]\Bigg\}\\
\tilde{W}^{(4)}(r,x)&=-\frac{g^2}{96 \tilde{M}^6 (r^2+a^2 x^2) r}\Bigg\{2 \tilde{M}^3 r^2-a^4 (r+\tilde{M}) x^2-a^2 \left[r \left(2 \tilde{M}^2+\tilde{M} r+r^2\right)-2 \tilde{M}^2 (r-\tilde{M}) x^2\right]\Bigg\}.
\end{aligned}
\end{equation}
With these in hand, we can compute all the relevant thermodynamic quantities,
\begin{equation}
\begin{aligned}
G M&=\frac{r_+^2+a^2}{2 r_+}+\frac{1}{2 r_+}\hat{Q}^2-\frac{a^2 g^2 r_+^5}{3 (r_+^2+a^2)^5} \hat{Q}^4+O(\hat{Q}^6)
\\
G J&= a G M=\frac{a(r_+^2+a^2)}{2 r_+}+\frac{a}{2 r_+}\hat{Q}^2-\frac{a^3 g^2 r_+^5}{3 (r_+^2+a^2)^5} \hat{Q}^4+O(\hat{Q}^6)
\\
G S& = \pi  (r_+^2+a^2)-\frac{\pi  a^2 g^2 r_+^2 (a^4+4 a^2 r_+^2+7 r_+^4)}{6 (r_+^2+a^2)^5} \hat{Q}^4+O(\hat{Q}^6)
\\
T&=\frac{r_+^2-a^2}{4 \pi  r_+ (r_+^2+a^2)}-\frac{1}{4 \pi  r_+ (r_+^2+a^2)}\hat{Q}^2-\frac{a^2 g^2 r_+ (a^2-r_+^2) (a^4+4 a^2 r_+^2+7 r_+^4)}{24 \pi (a^2+r_+^2)^7} \hat{Q}^4+O(\hat{Q}^6)
\\
\mu & = \frac{r_+}{r_+^2+a^2}\hat{Q}-\frac{a^2 g^2 r_+}{6 (r_+^2+a^2)^3}\hat{Q}^3+O(\hat{Q}^5)
\\
\Omega & = \frac{a}{a^2+r_+^2}+\frac{a g^2 r_+^2 (a^6+3 a^4 r_+^2+5 a^2 r_+^4-r_+^6)}{6 (r_+^2+a^2)^7} \hat{Q}^4+O(\hat{Q}^6),
\end{aligned}
\end{equation}
which can be checked to obey to the standard first law of black hole mechanics to $O(\hat{Q}^4)$. At extremality, where $T=0$, one finds
\begin{equation}
M^2=\frac{J^2}{G^2 M^2}+Q^2-\frac{g^2 G Q^4}{48 J}+O(Q^6).
\end{equation}

Finally, we discuss the axion field $\tilde{\varphi}^{(4)}(r,x)$, which can be shown to be given by
\begin{equation}
\tilde{\varphi}^{(4)}(r,x)=\frac{x}{96 \tilde{M}^3 (r^2+a^2 x^2)^2} \left\{a g r \left[8 \tilde{M}+(4+g^2) r\right]+a^3 g (4+g^2) x^2\right\}+\sum_{p=0}^{\infty}P_{2p+1}(x)S_{2p+1}(r),
\end{equation}
where $P_{\ell}(x)$ are Legendre polynomials of order $\ell$ and
\begin{equation}
\begin{aligned}
&\frac{{\rm d}}{{\rm d} r}\left[\Delta(r)\frac{{\rm d} S_{2p+1}(r)}{{\rm d} r}\right]-2 (p +1) (2 p +1)S_{2p+1}(r)=
\\
&\qquad\qquad\qquad\qquad \frac{(-1)^p a g (4-g^2) \sqrt{\pi } (4 p+3) \Gamma (p+1) \Gamma (2 p+4)}{3 \tilde{M} r^4 2^{2 p+5}}\left(\frac{a}{r}\right)^{2 p}\, {}_3\widetilde{F}_2\left[\begin{array}{c} p+1,p+2,p+\frac{5}{2} \\ p,2 p+\frac{5}{2}\end{array};-\frac{a^2}{r^2}\right] ,
\end{aligned}
\end{equation}
where ${}_3 \widetilde F_2$ is the regularized generalized hypergeometric function.
One can solve the above equation for each value of $p$ and for arbitrary values of $\tilde{M}$ and $a$. However, we are primarily interested in the case with $p=0$ and at extremality. For the purposes of the above equation, this means taking $\tilde{M}=r_+$ and $a=r_+$. For these particular values, one finds
\begin{multline}
S_1(r)=\frac{g (4-g^2)}{64 (r-r_+)^2 r_+^5 (r^2+r_+^2)}\Bigg\{2 (r-r_+) r_+^2 (2 r^2-r r_++r_+^2)+\pi  (r^2+r_+^2) (2 r^3-6 r^2 r_++6 r r_+^2-r_+^3)
\\
-2 \left(r^2+r_+^2\right) \Bigg[2 r (r^2-3 r r_++3 r_+^2) \arctan(r/r_+)-2 (r-r_+)^3 \log \Bigg(\frac{r-r_+}{\sqrt{r^2+r_+^2}}\Bigg)\Bigg]\Bigg\}.
\end{multline}
Expanding around $r=r_+$ reveals that
\begin{equation}
S_1(r)\approx \frac{g \left(4-g^2\right)}{32 r_+^4}+\frac{g (4-g^2)}{192 r_+^4}\left[12 \log \left(\frac{r}{r_+}-1\right)-1+3 \pi -6 \log 2\right] \left(\frac{r}{r_+}-1\right)+O\left[\left(\frac{r}{r_+}-1\right)^2\right],
\end{equation}
which shows that the axion is not continuously differentiable across the horizon and thus that the solution is singular. We expect this singularity to backreact on the metric and lead to tidal divergences of the sort explained in the main text. Indeed, we have confirmed this to be the case using the numerical methods described above.
\end{document}